\documentclass[prd,preprint,nofootinbib,superscriptaddress
  ,secnumarabic,showpacs,showkeys,
,amssymb, amsmath,mathrsfs,nobibnotes,aps,11pt]{revtex4}
\usepackage{graphicx,amsmath,amsfonts,amssymb,dcolumn,epsfig,bm,color}
\usepackage{graphicx, multirow}
\usepackage{dcolumn}
\usepackage{bm}
\usepackage{capt-of}
\usepackage{enumerate}
\usepackage{epsfig} 
\usepackage{wasysym} 
\usepackage{graphicx} 
\usepackage{amsfonts} 
\usepackage{amsbsy} 
\usepackage{amscd} 
\usepackage{hyperref}
\usepackage{pstricks} 
\usepackage{multirow}
\usepackage{ulem}
\usepackage{slashbox}
\makeatletter

\newcommand{\fmslash}[2][0mu]{
  \mathchoice
    {\fmsl@sh\displaystyle{#1}{#2}}
    {\fmsl@sh\textstyle{#1}{#2}}
    {\fmsl@sh\scriptstyle{#1}{#2}}
    {\fmsl@sh\scriptscriptstyle{#1}{#2}}}
\newcommand{\fmsl@sh}[3]{%
  \m@th\ooalign{$\hfil#1\mkern#2/\hfil$\crcr$#1#3$}}
\makeatother

\newcommand{\lsim}{{\;\raise0.3ex\hbox{$<$\kern-0.75em\raise-1.1ex\hbox{$\sim$}}\;}}
\newcommand{\gsim}{{\;\raise0.3ex\hbox{$>$\kern-0.75em\raise-1.1ex\hbox{$\sim$}}\;}}

\def\gs{\mathrel{
   \rlap{\raise 0.511ex \hbox{$>$}}{\lower 0.511ex \hbox{$\sim$}}}}
\def\ls{\mathrel{
   \rlap{\raise 0.511ex \hbox{$<$}}{\lower 0.511ex \hbox{$\sim$}}}}

\newcommand{\bad}{\begin{array}{ccc}}
\newcommand{\MET}{{{\fmslash E}_T}}

\begin{document}
\title{Looking for hints of a reconstructible seesaw model
\\at the Large Hadron Collider}
  
\author{Gulab Bambhaniya}
\email{gulab@prl.res.in}
\affiliation{Physical Research Laboratory (PRL), Ahmedabad-380009, Gujarat, India}
\author{Srubabati Goswami}
\email{sruba@prl.res.in}
\affiliation{Physical Research Laboratory (PRL), Ahmedabad-380009, Gujarat, India}
\author{Subrata Khan}
\email{subrata@prl.res.in}
\affiliation{Physical Research Laboratory (PRL), Ahmedabad-380009, Gujarat, India}
\author{Partha Konar}
\email{konar@prl.res.in}
\affiliation{Physical Research Laboratory (PRL), Ahmedabad-380009, Gujarat, India}
\author{Tanmoy Mondal}
\email{tanmoym@prl.res.in}
\affiliation{Physical Research Laboratory (PRL), Ahmedabad-380009, Gujarat, India}
\affiliation{Indian Institute of Technology, Gandhinagar, India.}

\date{\today}

\begin{abstract}
We study the production of heavy neutrinos at the Large Hadron Collider (LHC)
through the dominant s-channel production mode as well as the
vector boson fusion (VBF) process. We consider the TeV scale minimal linear seesaw
model containing two heavy singlets with opposite lepton number.
This model is fully reconstructible from oscillation data apart from an overall normalization constant which can  
be constrained from meta-stability of the electroweak vacuum and bounds coming from 
lepton flavor violation (LFV) searches. Dirac nature of heavy neutrinos in this model
implies suppression of the conventional same-sign-dilepton
signal at the LHC. We analyze the collider signatures with tri-lepton final state and
missing transverse energy as well as VBF type signals which are characterized by two
additional forward tagged jets. Our investigation reveals that due to  stringent 
constraints on light-heavy mixing coming from LFV and meta-stability bounds,
the model can be explored  only for light to moderate mass range of heavy neutrinos. 
We also note that in case of a positive signal, flavor 
counting of the final tri-lepton channel can give information
about the mass hierarchy of the light neutrinos.
\end{abstract}



\maketitle

\section{Introduction}

The discovery of the Higgs boson at the Large Hadron Collider both by 
ATLAS~\cite{:2012gk} and
CMS~\cite{:2012gu} collaborations have put the Standard Model (SM) on a firm
footing. However, no signal of physics beyond the Standard Model (BSM)
has been found so far at the LHC. On the other hand, 
convincing indications of BSM
physics have already emerged from the phenomenon of neutrino oscillation
observed in terrestrial experiments.  These results have conclusively
established that neutrinos have non-zero mass and flavor mixing.
Oscillation data together with the cosmological bound on sum of
neutrino masses ($\Sigma m_i < 0.23$ eV including the PLANCK data
~\cite{planck}) indicate that neutrino masses are much smaller as
compared to the other fermions in the SM.  Such small masses can be
generated naturally by the seesaw mechanism.  The origin of seesaw is
the dimension 5 effective operator $\frac{c_5}{M} L L HH$, where $L$($H$)
being the SM lepton(Higgs) doublet and $c_5$ is a
dimensionless coupling, $M$ is the mass scale at which the effective operator gets generated~\cite{Weinberg:1979sa}.   
Such operators arise by integrating out heavy
fields added to the SM Lagrangian and they violate lepton number by two
units.  The smallness of neutrino mass in these models is related
to the scale of lepton number violation which is required to be very high
$\sim \mathcal{O}( 10^{15} \textrm{ GeV})$ to generate neutrino masses in the right ballpark. The most
economical in terms of particle contents is the type-I seesaw in which
heavy singlet right-handed neutrinos are added to the 
SM Lagrangian~\cite{Minkowski:1977sc,Yanagida:1979as,GellMann:1980vs,Glashow:1979nm,Mohapatra:1979ia}. 
However, the natural seesaw scale is far beyond the reach of the LHC.
To have signatures of seesaw models at the LHC,  the heavy neutrino $(N)$ mass needs to be $\sim \mathcal{O}$ (TeV). 
However, if one lowers the scale of seesaw to TeV then it also
requires much smaller neutrino Yukawa couplings ($\sim 10^{-6}$) to
obtain correct light neutrino masses. Such small Yukawa couplings lead
to suppression of the production of the heavy neutrinos in natural TeV
scale Type-I seesaw models.  This leads to the question whether it is
possible to achieve both the requirements simultaneously, i.e. having
TeV scale heavy neutrinos along with large Yukawa coupling leading to large 
light-heavy mixing.  Such possibilities can be realized in some specific mass textures
~\cite{Pilaftsis:1991ug,Gluza:2002vs,Pilaftsis:2004xx,Kersten:2007vk,Xing:2009in,He:2009ua,Ibarra:2010xw,
  Deppisch:2010fr,Adhikari:2010yt}.  Other options include models with
higher-dimensional operators arising due to exchange of new particles
belonging to larger representations
~\cite{Gogoladze:2008wz,Babu:2009aq,Bonnet:2009ej,Kanemura:2010bq,Liao:2010rx,Picek:2009is,Liao:2010cc,Kumericki:2012bh,McDonald:2013kca}, radiative mass
generations~\cite{Zee:1980ai,Zee:1985id,Babu:1988ki,Krauss:2002px,Ma:2006km,Aoki:2008av,Dev:2012sg,Gustafsson:2014vpa} etc.  
One of the most popular options to generate TeV
scale seesaw is through the inverse seesaw models in which one
includes additional singlet states.  These models were first proposed
in the context of E(6) Grand Unified Theories~\cite{Mohapatra:1986bd}. 
In these
models the seesaw scale is decoupled from the scale of lepton number
violation and the smallness of neutrino mass originates from the
small lepton number violating terms in the Lagrangian.

In type-I Seesaw model the heavy and light neutrinos are both Majorana
particles. It is well known that Majorana nature of neutrinos can be
established by observing a positive signal in neutrino-less double beta
decay experiments. 
It was  noticed in 
~\cite{Keung:1983uu}, in 
the context of the Left-Right symmetric model 
that resonant production of $N$ and its subsequent decay giving same-sign
di-lepton (SSDL) signal in colliders 
can also constitute evidence for Majorana nature of neutrinos. 
Given the importance
of this signal, there have been several studies of this channel
at the hadron colliders
~\cite{Datta:1993nm, Han:2006ip,Bray:2007ru,delAguila:2007em,
delAguila:2008cj,Atre:2009rg}  
including searches at the LHC~\cite{Chatrchyan:2012fla}. 
Enhanced contribution from infrared t-channel, 
especially for heavier masses was 
proposed~\cite{Dev:2013wba,Das:2014jxa} together with s-channel production.

The heavy neutrinos in inverse seesaw model are of pseudo-Dirac nature
and in this case the SSDL signal is suppressed by the small lepton
number violating coupling.  For such models the heavy neutrinos are
produced by the s-channel process along with a charged lepton. This
neutrino further decays to a second lepton (of sign opposite to the first
lepton to conserve lepton number) together with a W-boson. The W-boson can further
decay leptonically to produce a lepton and a neutrino. Thus the final signal 
consists of tri-lepton and missing energy which is expected to have tiny contamination 
from standard model backgrounds.  Detailed studies including the SM
background in the context of pseudo-Dirac neutrinos have been done in
~\cite{delAguila:2008cj,delAguila:2008hw}. 
Similar studies in the context of Left-Right
symmetric model, non-minimal supersymmetric inverse seesaw models and Type-III 
seesaw model have been performed in~\cite{Chen:2011hc},~\cite{Das:2012ze} and~\cite{Eboli:2011ia} 
respectively. Experimental searches for multi-lepton signals have
been carried out by the CMS collaboration using an integrated
luminosity of 19.5 $fb^{-1}$ with center of mass energy $\sqrt{s} = 8$ 
TeV at the LHC~\cite{Chatrchyan:2014aea}. They considered at least three
leptons in the final state using a search strategy not specific to any
particular model.

In this work, we consider the minimal linear seesaw model (MLSM)
studied in~\cite{Gavela:2009cd,Khan:2012zw} as an example of the TeV scale
seesaw model. This is a variant of the inverse seesaw model but in this case
the minimal scheme consists of adding just two heavy singlets with
opposite lepton number as opposed to four heavy neutrinos in canonical minimal
inverse seesaw models~\cite{Malinsky:2009df}. It was shown in~\cite{Gavela:2009cd} that the Yukawa
couplings matrices for this model can be fully reconstructed in terms
of the oscillation parameters apart from an overall normalization
factor. It was further shown in~\cite{Khan:2012zw} that this normalization
constant can be constrained from consideration of the
meta-stability of the electroweak vacuum and lepton flavor
violation bounds.  The heavy neutrinos in this model are of Dirac type 
and the SSDL signal is suppressed\footnote{ Due to same reason 
heavy neutrino contribution towards $0\nu\beta\beta$ is suppressed~\cite{Khan:2012zw}.}. In the context of this model we consider two possible
production channels for the heavy neutrinos resulting in two different
classes of signals.  The first one of this is the s-channel 
process to produce heavy Dirac neutrinos
associated with a lepton and finally 
giving the tri-lepton and missing energy signal.  The second
one is the production of heavy neutrinos through vector boson fusion (VBF)  
in which two electroweak vector bosons coming from two partons 
`fuse' to produce the signal under consideration (tri-leptons) along with two 
highly forward jets. It becomes important in the context of hadron colliders 
since the tagging of forward jets allows us to reduce the background considerably. 
Also the lack of color exchange between these jets makes the central region 
free from the color activities and this is exploited by vetoing central jets;
see ~\cite{Rainwater:1999gg} and references therein in the context of Higgs search. 
This helps in minimizing the backgrounds further. For these reasons VBF remains an important channel to look for 
new physics~\cite{Datta:2001hv,Choudhury:2003hq,Cho:2006sx} at hadron colliders. 

We consider
both normal hierarchy (NH) as well as inverted
hierarchy (IH) for the light neutrino mass spectra. 
We also estimate the corresponding standard model 
backgrounds for the 14 TeV LHC. In each case, we perform a realistic 
simulation with extensive event selections using {\tt MadGraph} and 
{\tt PYTHIA}.

The paper is organized as follows: Sec.~\ref{sec:model} contains a
brief description of the model. The production and
decay of the right handed neutrino at LHC, are discussed in
Sec.~\ref{sec:phenomenology}. Simulation details and results are
presented in Sec.~\ref{sec:simulation_result}, while in
Sec.~\ref{sec:discovery_pot} we discuss discovery potential of the
signals at the LHC.  Finally, we conclude in Sec.~\ref{sec:conclusion}.

\section{The linear seesaw model}\label{sec:model}



The most general Lagrangian containing heavy singlet fields $N_R$ and
S with opposite lepton numbers, is given by
\begin{eqnarray}
-{\cal{L}} = \overline{N}_R Y_{\nu} \tilde{\phi}^{\dag} l_{L} 
	    +  \overline{S} Y_{S} \tilde{\phi}^{\dag} l_{L}
	    +  \overline{S}M_N N_R^c + \frac{1}{2}\overline{S} \mu S^c 
	    + \frac{1}{2} \overline{N_R} {\mu_N} N_R^c 
	    + {\mathrm h.c.}, 
\label{lag:full}
\end{eqnarray}
where $l_L = (\nu_x,x)_L^T$, $x = {e, \mu, \tau}$.

Once the symmetry is broken spontaneously, the Higgs field $\phi$ obtains a 
vacuum expectation value (VEV) equal to $v/\sqrt{2}$.  This generates the
Dirac mass term $m_D = Y_\nu v/\sqrt{2} $ and the lepton number
breaking mass term $m_S = Y_{S} v/\sqrt{2} $. In the linear seesaw models
~\cite{Gu:2010xc,Zhang:2009ac,Hirsch:2009mx} one assumes $m_S$ to be
small and non-zero while the $\mu$ and the $\mu_N$ terms are set to zero. 
This can be done since they contribute towards light neutrino mass in the sub-leading orders~\cite{Bazzocchi:2010dt}. 
Since lepton number violating mass
 terms are set to zero, the heavy neutrinos are purely Dirac type. 
Then the mass matrix 
takes the form
\begin{equation} 
\mathcal{M}_{\nu} =
\bad
\begin{pmatrix}
0 & m_D^T &  m_S^T \\
m_D & 0 & M_N \\
m_S & M_N^T & 0 \\
\end{pmatrix},
\end{array}
\label{eq:linearseesaw}
\end{equation} 
in the $(\nu_L, N_R^c, S^c)$ basis. 

The minimal model which can successfully generate two light neutrinos 
with non-zero mass is when only two extra heavy singlets 
are added to the SM Lagrangian. This is called the Minimal Linear Seesaw Model 
(MLSM)~\cite{Gavela:2009cd,Khan:2012zw}. The full mass matrix has 
dimension $5 \times 5$  and can be written as ,

\begin{equation}
\mathcal{M}_\nu = 
\left(\begin{array}{cc} 
0 & {m_D^\prime}^T \\
{m_D^\prime} &  M  
\end{array}\right),
\label{mrecast}
\end{equation}
where $m_D^{\prime T} = (m_D^T,m_S^T)$ and 
\begin{equation}
M = 
\left(\begin{array}{cc} 
0 & M_N \\
M_N & 0 
\end{array}\right). 
\label{msinglet} 
\end{equation}
For the minimal case $M_N$ is just a number, not a matrix. 
$\mathcal{M}_\nu$ can be diagonalized
by a 5 $\times$ 5 unitary matrix $U_0$ as
\begin{eqnarray}
\label{diagonal}
U_0^T \mathcal{M}_\nu\, U_0 = \mathcal{M}_\nu^{\mathrm diag},
\end{eqnarray}
where $\mathcal{M}_\nu^{\mathrm diag} 
=\mbox{diag}(m_1\,,m_2\,,m_3\,,M_1\,,M_2)$. 
Following a two-step diagonalization procedure~\cite{Grimus:2000vj},
$U_0$  can be expressed as,
\begin{eqnarray}
U_0=  
\left(\begin{array}{cc}  
\left(1-\frac{1}{2}\,\epsilon \right)U_{\nu} & {m_D^{\dagger}} (M^{-1})^{\ast}U_{R}\\
-M^{-1}m_D\, U_{\nu} & \left(1-\frac{1}{2}\,\epsilon'\right)U_{R}
\end{array} \right) 
 \equiv \left(\begin{array}{cc}
U_{L} & V\\
S & U_{H }
 \end{array} \right)
\label{bdmatrix},  
\end{eqnarray} 
where, $U_L$ is the 
$U_{PMNS}$
mixing matrix,  
and $V$, $S$ are  the light-heavy mixing matrices.
Interaction of heavy neutrinos with the SM fields are determined
by the mixing matrix V, whose elements will be denoted as $V_{lN}$ hereafter.
We would notice afterwards that the strong constraints on some elements of this matrix {\it i.e.} 
$V_{e N}$ and $V_{\mu N}$ would restrict the production signal. 
The diagonalizing matrix is now non-unitary  which is characterized by the factor  $(1 - \epsilon/2)$. 
The non-unitary corrections $\epsilon$ and $\epsilon'$ are given in \cite{Grimus:2000vj,Khan:2012kc}.
$U_\nu$ is the unitary component of $U_{PMNS}$ which is same as 
$U_{PMNS}$ for $\epsilon << 1$. We use the standard parametrization for this: 
\begin{equation}
U_\nu   =  \left(
 \begin{array}{ccc}
 c_{12} \, c_{13} & s_{12}\, c_{13} & s_{13}\, e^{-i \delta}\\
 -c_{23}\, s_{12}-s_{23}\, s_{13}\, c_{12}\, e^{i \delta} &
 c_{23}\, c_{12}-s_{23}\, s_{13}\, s_{12}\,
e^{i \delta} & s_{23}\, c_{13}\\
 s_{23}\, s_{12}-\, c_{23}\, s_{13}\, c_{12}\, e^{i \delta} &
 -s_{23}\, c_{12}-c_{23}\, s_{13}\, s_{12}\,
e^{i \delta} & c_{23}\, c_{13}
 \end{array}
 \right) P \, ,
\label{upmns_param}
\end{equation}
where 
$c_{ij} = \cos \theta_{ij}$, $s_{ij} = \sin \theta_{ij}$ and 
$\delta$ denotes the Dirac CP phase. The Majorana phase matrix $P$ is 
expressed as   
$P = {\rm diag}( e^{-i \alpha}, e^{i \alpha},1)$, there is only one 
Majorana phase because  one of the mass eigenvalues is zero.  
In table \ref{table:oscillation_param}, we have  presented 
the 3$\sigma$ allowed range of oscillation parameters.  
Note that the phases are completely unconstrained at present. 

Using the seesaw approximation  one obtains  the light neutrino mass matrix, 
\begin{equation} 
m_{light}= m_D^{\prime T} M^{-1} m_D^{\prime}. 
\label{mlight} 
\end{equation} 
This being a rank 2 matrix the light neutrinos belonging to this 
model are hierarchical. Thus there are two possible mass spectra:   
\begin{itemize} 
\item 
Normal Hierarchy (NH): ~~~  $(m_1 < m_2 < m_3)$
\item 
Inverted Hierarchy (IH): ~~~ $(m_3 << m_2 \approx m_1 ).$    
\end{itemize} 
 
In MLSM, $Y_{\nu}$ and $Y_{S}$ are $3 \times 1$ matrices (cf. Eq.~\ref{lag:full}) and can be considered 
as two independent vectors  
\begin{equation} 
Y_\nu  \equiv  y_\nu {\hat{\bf{a}}};~~
Y_{S} \equiv y_s {\hat{\bf{b}}} ,
\label{vector} 
\end{equation} 
where ${\hat{\bf{a}}}$ and ${\hat{\bf{b}}}$ denotes complex vectors with  
unit norm  while 
$y_{\nu}$ and $y_{s}$ represent
the norms of the Yukawa matrices $Y_{\nu}$ and $Y_{S}$, respectively.
Using  Eq.~\ref{mlight}  and \ref{vector} 
one can reconstruct the Yukawa matrices 
$Y_\nu$ and $Y_S$ in terms of the oscillation parameters 
barring an overall normalization factor. 
The parametrization of the Yukawa matrices depend on the mass hierarchy 
and can be expressed 
as~\cite{Gavela:2009cd, Khan:2012zw},
\begin{eqnarray}\label{eq:yukawa}
Y_{\nu} &=& \frac{y_{\nu}}{\sqrt{2}}\left(\sqrt{1+\rho}~U_j^{\dag} + e^{i\frac{\pi}{2}}\sqrt{1-\rho}~U_k^{\dag}\right), \nonumber \\
Y_{S} &=& \frac{y_{s}}{\sqrt{2}}\left(\sqrt{1+\rho}~U_j^{\dag} - e^{i\frac{\pi}{2}}\sqrt{1-\rho}~U_k^{\dag}\right),
\label{pmtz-nh} 
\end{eqnarray}
where, $j =2\,, k=3$ for NH and $j=2\,,k=1$ for IH. 
$U_j$'s denote the columns of the unitary matrix $U_{\nu}$ that diagonalizes
the light neutrino mass matrix $m_{light}$ in Eq.~\ref{mlight}.  
The parameter $\rho$ is given as,  
\begin{equation}\label{eq:rho}
\rho=\frac{\sqrt{1+r}-\sqrt{r}}{\sqrt{1+r}+\sqrt{r}} \;\;(NH), \, \quad 
\rho=\frac{\sqrt{1+r}-1}{\sqrt{1+r}+1} \;\; (IH).  
\end{equation}
Here $r$ denotes the ratio of the solar and atmospheric mass  squared differences, 
$r={\Delta m_{\odot}^2}/{\Delta m^2_{atm}}$, with  $\Delta m_{\odot}^2 \equiv m_2^2 - m_1^2$ and 
$\Delta m^2_{atm} \simeq m_3^2 - m_1^2 \,~~\left(m_2^2 - m_3^2\right)$ for NH (IH). 

\begin{figure}[t]
\includegraphics[width=7.0cm, angle =0]{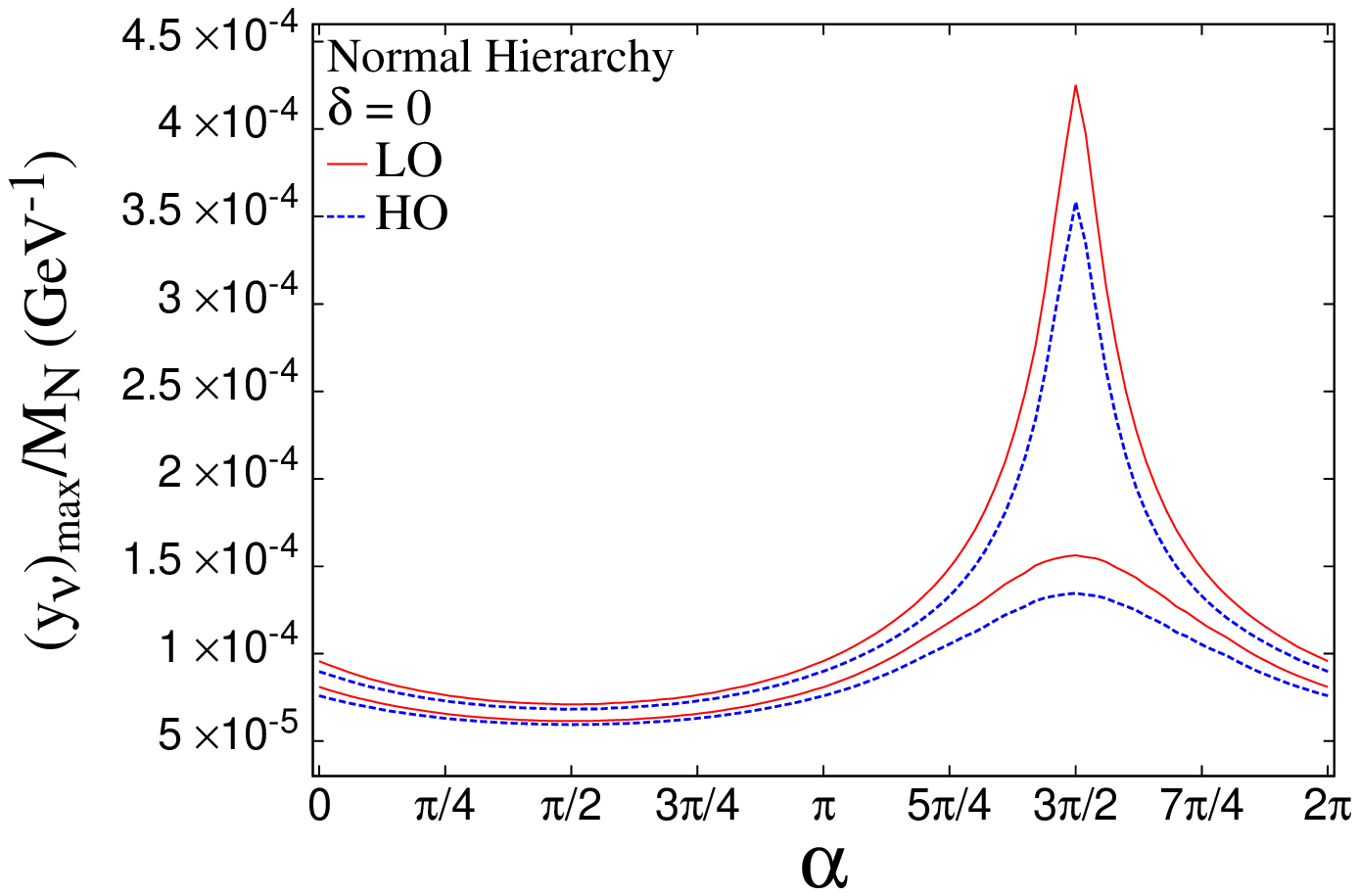} 
\includegraphics[width=7.0cm, angle =0]{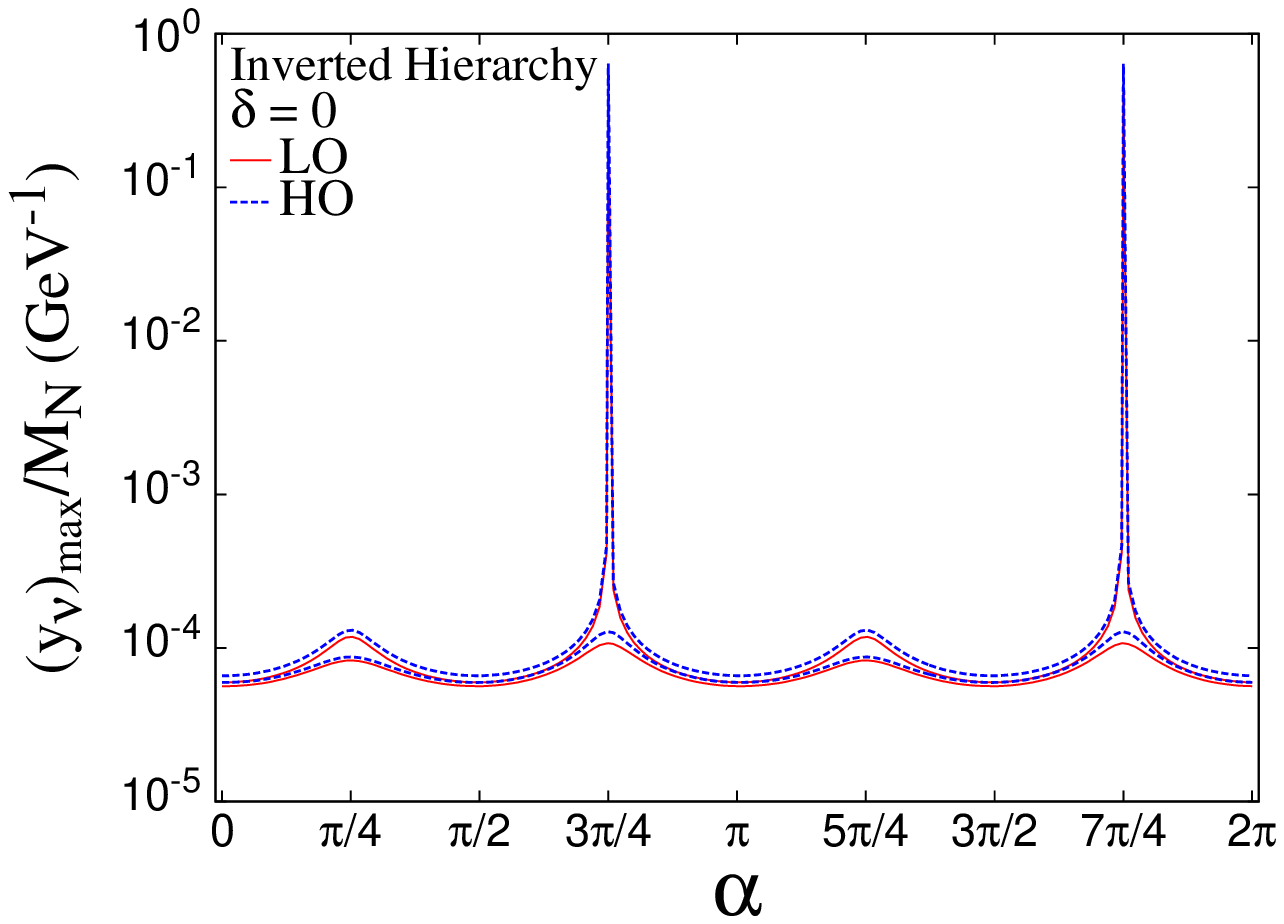} 
\caption{Bound on $y_{\nu}/M_N$ as a function of Majorana phase $\alpha$, varying the oscillation parameters in the allowed $3\sigma$ range.
Red-solid(Blue-dashed) curve corresponds to atmoshpheric angle ($\theta_{23}$) residing in LO(HO) region.
(Left plot) The plot is for NH scenario, where highest allowed value of $y_{\nu}/M_N$ lies in LO region. 
(Right plot) The same plot for IH scenario. 
}
 \label{fig:ynumr_vs_alpha}  
\end{figure}

The overall coupling $y_{\nu}$ can be constrained from the metastability of 
the electro-weak vacuum and LFV ~\cite{Khan:2012zw}. 
For NH the most stringent constraint comes from LFV, whereas for 
IH case vacuum meta-stability constraint
is more restrictive.
This is because of cancellations occurring for IH for LFV processes 
\cite{Khan:2012zw}. 
The dependence of the bound on $y_\nu$ from meta-stability and LFV 
on the heavy neutrino mass has been shown in \cite{Khan:2012zw}. 
The metastability bound  on $y_\nu$   
varies approximately in the range
0.4 - 0.5 for $M_N$ varying in the range 100 - 1000 GeV. 
This bound is indpendent of the oscillation parameters. 
However, significant variation on the bound on $y_\nu$ 
from LFV constraint is possible  within the allowed range of 
oscillation parameters, mostly due to unconstrained phases, 
$\delta$ and $\alpha$.  
Details of these dependence can be followed from Fig.~\ref{fig:ynumr_vs_alpha}. 
For a particular $M_N$, the strength of the signal at LHC would depend 
on the value of $y_\nu$.  To maximise the signal we therefore 
choose the value of $y_\nu$ at the peak for NH case. However, for IH case the peak value is much above the vacuum metastability bound and therefore we choose maximum allowed value of $y_{\nu}$ satisfying the
metastability bound.
The corresponding parameter 
values are depicted in table \ref{table:oscillation_param} for NH (case I) 
and IH. Note that, the above mentioned cancellations within the terms, ensure the peak position corresponds to $\alpha + \delta = 
3 \pi/2 (3\pi/4) $ for NH (IH), which is also evident in Fig.~\ref{fig:ynumr_vs_alpha}.    
We have chosen $\delta =0$ in our analysis. For some other values of 
$\delta$, the phase $\alpha$ has to be chosen so that one is at the peak. 
In Fig.~\ref{fig:ynumr_vs_alpha} we also show the variation of this bound 
with respect to  the  $\theta_{23}$ mixing 
angle in lower octant (LO, $\theta_{23} < \pi/4$) and higher octant 
(HO, $\theta_{23} > \pi/4$).  
The $y_\nu$ value 0.4(0.075) corresponds to IH(NH: Case-I) scenario for $M_N = 100$ GeV, which we will use in our analysis. 
%
These  will be translated into 
the bounds on the mixing matrix 
elements, $V_{lN}$, depending on the 
heavy neutrino mass $M_N$.  Since $y_s$ is extremely small ($\mathcal{O}(10^{-10})$), $Y_{S}$ does not play 
any role in determining $V_{lN}$. 
The elements of the matrix $V({\rm{or~}} V_{lN})$ can be expressed in terms of $U_{PMNS}$ matrix, $\rho$ and $y_\nu$ as follows:
\begin{eqnarray}
  V_{e{N_1}}  &=& \frac{-i}{\sqrt{2}M_N} \frac{y_\nu\,v}{2}
	  \left[\sqrt{1+\rho}\; (U_{PMNS})_{12}^*+i\,\sqrt{1-\rho}\;(U_{PMNS})_{11}^* \right] \nonumber\\
	   &\simeq& \frac{y_\nu\,v}{4M_N}  \left[e^{i(\alpha+\delta)}(-2+\sqrt{r})\,r^{\frac{1}{4}}\, s_{12}-2\, i\, s_{13}\right]
		  + \mathcal{O}\left((\sqrt{r},s_{13})^2\right)\nonumber\\
   V_{\mu{N_1}}  &=& \frac{-i}{\sqrt{2}M_N} \frac{y_\nu~v}{2} 
	  \left(\sqrt{1+\rho} (U_{PMNS})_{22}^*+i\sqrt{1-\rho}(U_{PMNS})_{21}^* \right) \nonumber \\
	  &\simeq& \frac{y_\nu\,v}{4M_N}\left[ (-2+\sqrt{r})\,(e^{i\alpha}\,r^{\frac{1}{4}}\,c_{12}\,c_{23}+i\,s_{23}) +
		  2\,e^{i\,(\alpha+\delta)}\,r^{\frac{1}{4}}\,s_{12}\,s_{23}\,s_{13}\right]
		  + \mathcal{O}\left((\sqrt{r},s_{13})^2\right). ~~~~~~~~~
\end{eqnarray}
The above expressions are for NH scenario and similar expressions can be
computed for IH also. The element $V_{e{N_2}}$
$\left(V_{\mu{N_2}}\right)$ differs from  
$V_{e{N_1}}$($V_{\mu{N_1}}$) by a phase factor. 
Note that 
in Table ~\ref{table:oscillation_param},   
we also consider a second set of oscillation parameters for NH (NH: Case II) corresponding 
to a lower value of $y_\nu$ of 0.056 with $\theta_{23}$ in the higher octant.
This value is chosen such  that   
$V_{\mu N}$ is maximum and muon signal may be larger, since muon has higher efficiency for detection.

\begin{table}
\begin{tabular}{|c||c|c|c|c|c|c|}
\hline
\backslashbox{Bound~\hskip -0pt}{\hskip -0pt ~Parameter~}
&$\Delta_{\odot}^2[10^{-5}\,\textrm{eV}^2]$ & $\Delta_{\textrm{atm}}^2[10^{-3}\,\textrm{eV}^2]$ & $\sin^2\theta_{12}$ & $\sin^2\theta_{23}$ & $\sin^2\theta_{13}$ & $\delta$  \\
\hline\hline
3~$\sigma$ range (NH)~&~ $7.12-8.20 $ ~&~ $2.31-2.74$ ~&~ $0.27-0.37$ ~&~ $0.36-0.68$ ~&~ $0.017-0.033$ ~&~ $0-2\pi$ ~ \\
~\hskip 45pt(IH)&~&~$2.21-2.64$~&~~&~$0.37-0.67$~&~~&~\\
\hline
Used value (NH: Case - I) ~&~ $7.15$ ~&~ $2.73$ ~&~ $0.27$ ~&~ $0.36$ ~&~ $0.033$ ~&~ $0.0$ \\
\hline
Used value (NH: Case - II) ~&~ $7.13$ ~&~ $2.73$ ~&~ $0.27$ ~&~ $0.68$ ~&~ $0.033$ ~&~ $0.0$ \\
\hline
Used value (IH) ~&~ $7.25$ ~&~ $2.40$ ~&~ $0.34$ ~&~ $0.57$ ~&~ $0.021$ ~&~ $0.0$ \\
\hline
\end{tabular}
\caption{ Allowed 3$\sigma$ ranges of oscillation parameters and benchmark values of these parameters
used in our analysis to get the signal allowed by LFV and vacuum metastability. Case-I corresponds to 
the peak in Fig.~\ref{fig:ynumr_vs_alpha}(Left panel), while Case-II corresponds to a lower value of $y_\nu/M_N$ , for which  $V_{\mu N}$ is maximum.
 Value of  Majorana phase $\alpha$ is set at $3\pi/2 \, (3\pi/4)$ for NH(IH) scenario. }
\label{table:oscillation_param}
\end{table}

To get some perspective on the degree of suppression in cross section coming 
from these constraints we note down the corresponding $V_{lN}$ values for $M_N=100$ GeV 
 as: $V_{eN} = 1.95\times 10^{-3}$, $V_{\mu N} = 2.93\times 10^{-2}$ and $V_{\tau N} = 8.83\times 10^{-2}$ for NH (Case-I) scenario, whereas, 
 $V_{eN} = 1.43\times 10^{-3}$, $V_{\mu N} = 4.14\times 10^{-2}$ and $V_{\tau N} = 5.48\times 10^{-2}$ for NH (Case-II)  respectively. 
 For IH these values are $V_{eN} = 0.48$, $V_{\mu N} = 4.15\times 10^{-9}$ and $V_{\tau N} = 0.109$.
Note that since our model is fully reconstructible and the only unknown
parameter is $y_\nu$ which can be constrained from LFV and meta-stabilty 
bounds, we have definite predictions for the parameters $V_{lN}$ 
and these values are different for NH and IH scenarios.  
Bounds on $V_{lN}$ can also come from Electroweak Precision Data (EWPD)
\cite{Blas:2013ana}. 
Our bounds for NH are consistent with these bounds. 
For IH we get a larger value for $V_{eN}$. However it is to be 
noted that the EWPD   
bounds are obtained assuming mixing with a single charged lepton  
and can be evaded in presence of cancellations or mixing with the 
other charged leptons \cite{delAguila:2008cj}.

\begin{figure}[t]
\includegraphics[width=7.0cm, angle =0]{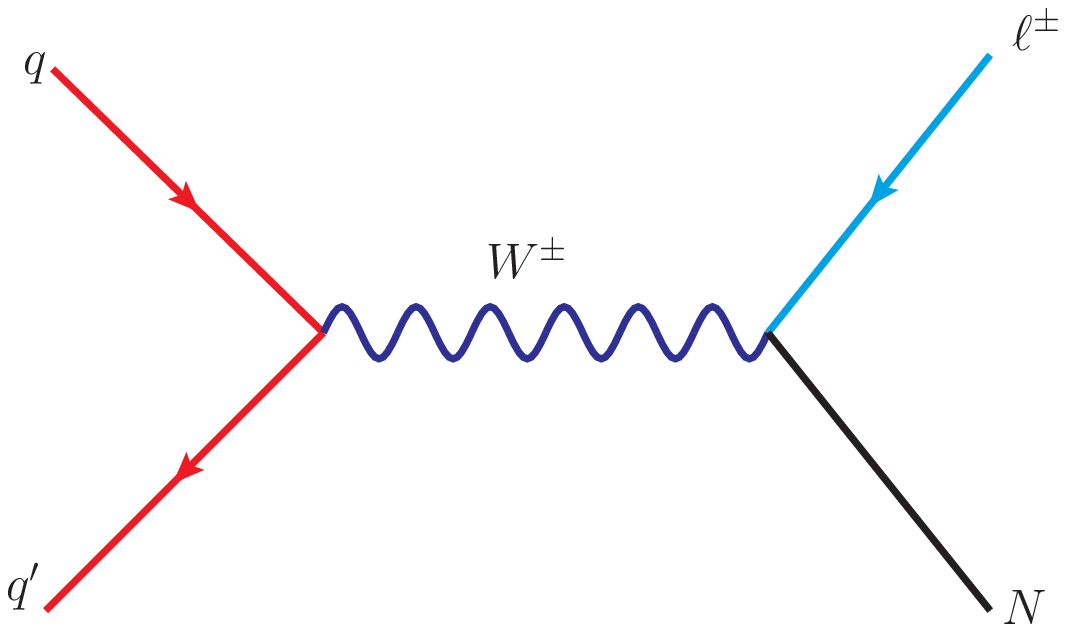} 
\includegraphics[width=6.0cm, angle =0]{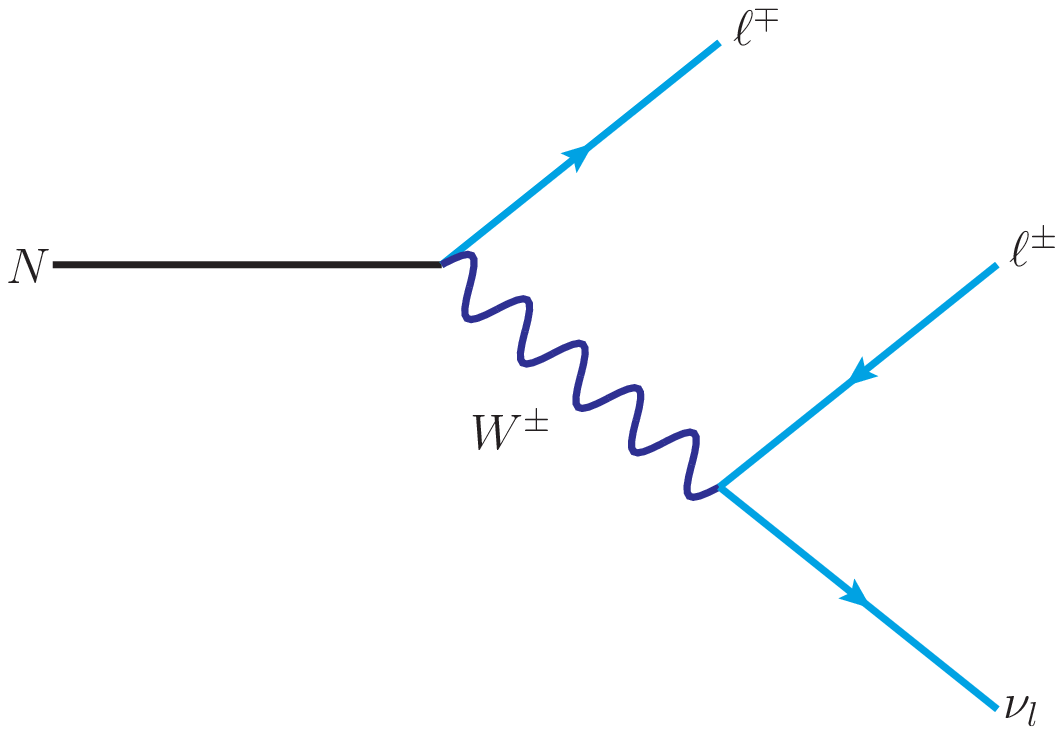} 
\caption{(Left plot) Leading order s-channel diagram for heavy neutrino production at hadron colliders, 
and (Right plot) representative diagram for one of the decay mode of the heavy neutrino. These two figures lead to 
tri-lepton $+ ~\MET$ signal considered in the analysis.}
 \label{fig:pp_to_Rnl}  
\end{figure}

\section{Phenomenology at the LHC}\label{sec:phenomenology}

\begin{figure}[t]
\includegraphics[width=7.0cm, angle =0]{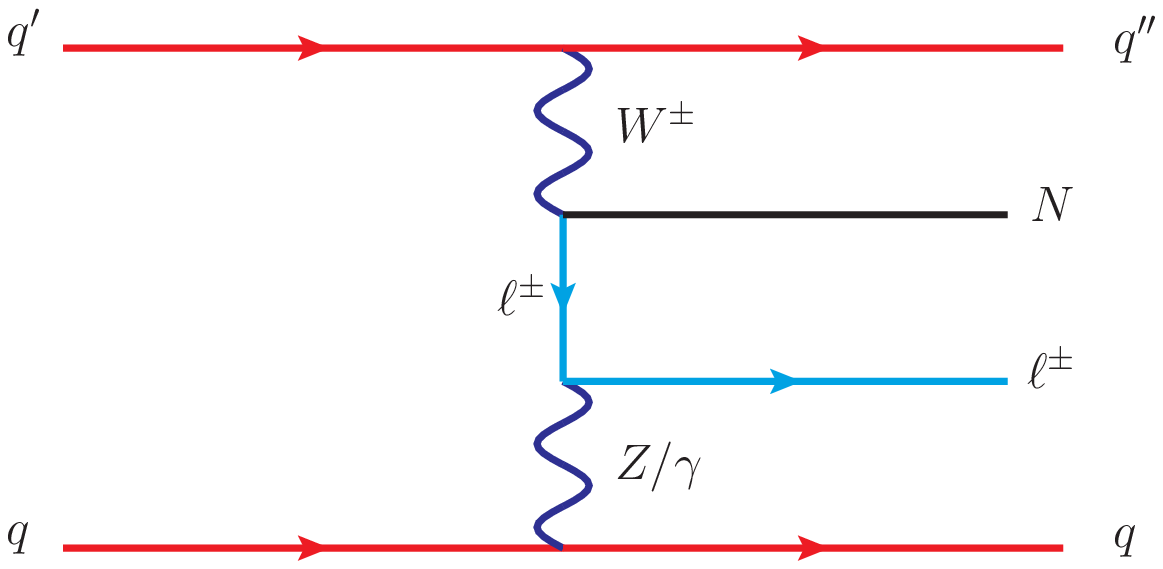}
\includegraphics[width=7.0cm, angle =0]{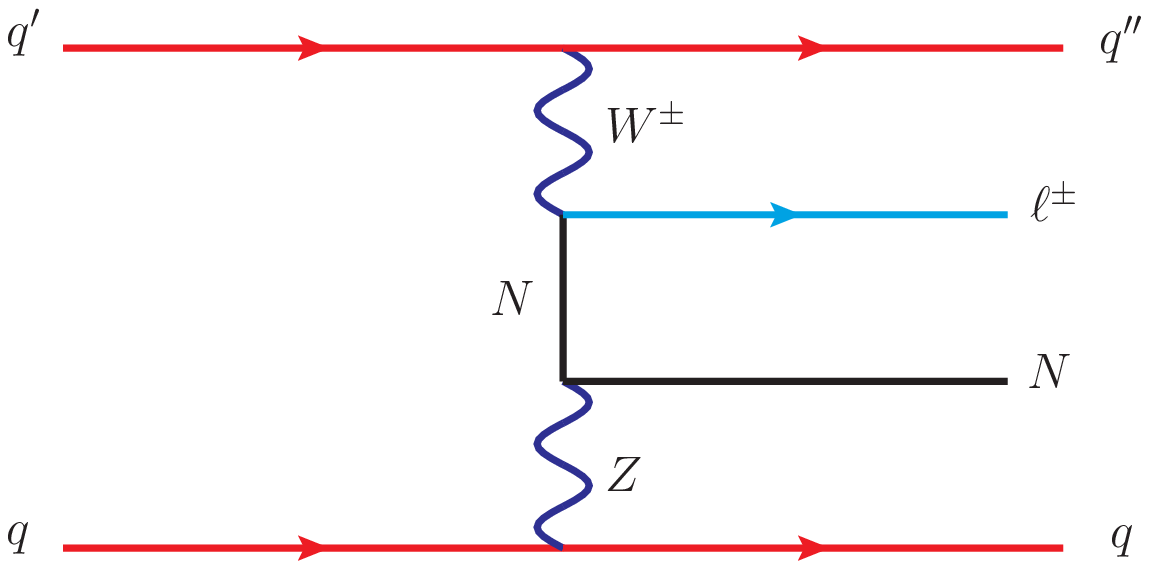}
\includegraphics[width=7.0cm, angle =0]{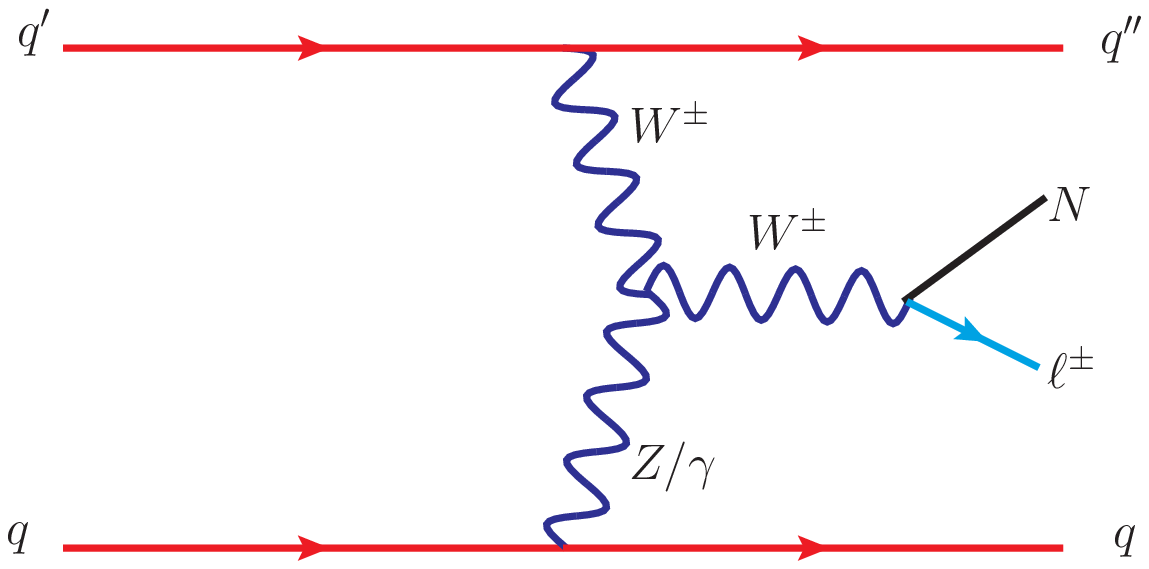}
\includegraphics[width=7.0cm, angle =0]{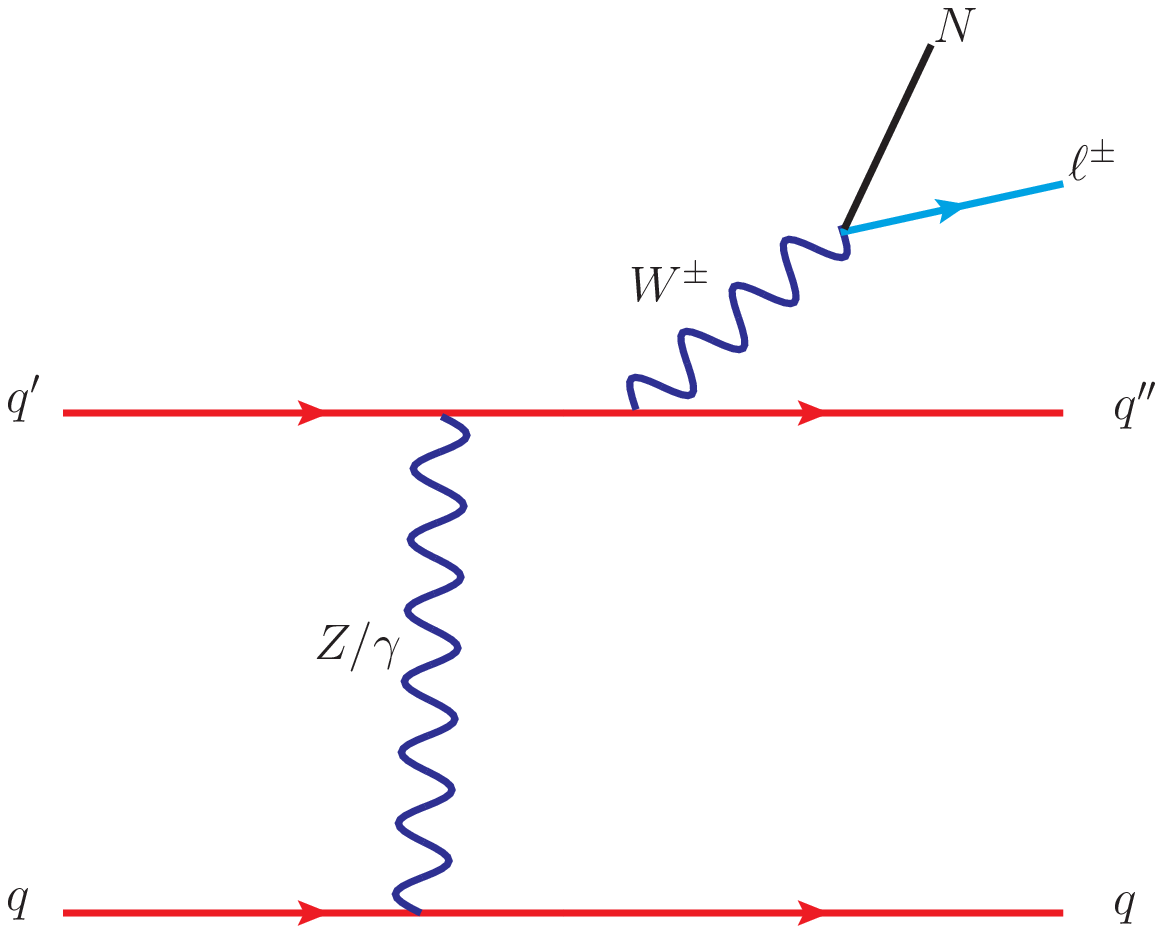}
\caption{Representative parton level diagrams contributing to $N \ell j j$ production through 
vector boson fusion at hadron colliders. Mirror diagrams are not shown here and also the last 
diagram is one of the four diagrams 
with $W^{\pm}$ emitting from each of the quark legs.}
 \label{fig:pp_to_Rnljj}
\end{figure}

\begin{figure}[t]
\includegraphics[width=6.5cm, angle =-90]{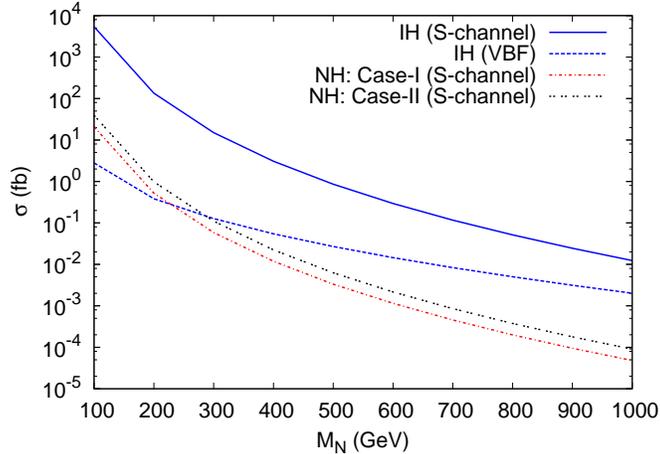} 
\caption{The total cross section is shown for production of heavy neutrino associated with light lepton ($p p \rightarrow N \ell$, where $\ell = e, \mu$ )
at the 14 TeV LHC through the leading order s-channel process
, while dotted lines represent VBF production cross section.
}
\label{fig:xsection_Nl_Nljj}
\end{figure}

The dominant production channel of the heavy neutrinos at LHC is the
s-channel process through virtual W-boson exchange. At the leading order 
the parton level process
($q \bar{q'} \to W^{\pm} \to \ell^{\pm} N$) is depicted in Fig.~\ref{fig:pp_to_Rnl}(left plot).
The heavy neutrinos can also be produced through the VBF
process where production of $N$ is associated with two forward jets. Fig.~\ref{fig:pp_to_Rnljj} 
contains the representative parton level Feynman diagrams for VBF processes\footnote{Note that there are some
diagrams which are not truly VBF type, i.e. two gauge boson are not fused via t-channel 
({\sl e.g.} bottom right diagram in Fig.~\ref{fig:pp_to_Rnljj}), but they
can lead to the same final states. 
These diagrams are necessary for the requirements of 
gauge invariance and included both for BG~\cite{Bozzi:2007ur, Jager:2006cp} and signal calculations.}.
Estimated total production cross sections of these heavy Dirac neutrinos 
at the 14 TeV LHC in IH scenario are shown in Fig.~\ref{fig:xsection_Nl_Nljj} for both $s$-channel(solid-line) as well as VBF (dashed-line). 
For NH scenario the $s$-channel production crossections are shown in the 
same figure for two different cases ({\it c.f.} Table~\ref{table:oscillation_param}), Case-I ( Red dot-dashed line)
and Case-II ( Black double dotted line).
Basic cuts such as 
${p_T}_{\ell} > 20$ GeV and $|\eta_{\ell}| < 2.5$
 are applied and $y_\nu$ values 
mentioned in the previous section are used. 
It is seen from the figure that although case II corresponds to 
a lower value of $y_\nu$ since $V_{\mu N}$ is larger, the production
cross-section is slightly larger. Since the VBF cross-section is much lower we 
do not present the VBF cross-section for the NH case. 
In these analyses CTEQ6L1~\cite{Pumplin:2002vw} parton
distribution functions have been used with 
the factorization scale set at the heavy neutrino mass $M_N$.

Heavy neutrinos $N$ can decay into charged lepton or neutrino associated with gauge (or Higgs) boson. 
\begin{equation}
N \rightarrow W^{\pm} l^{\mp} / Z \nu_l / H \nu_l,
~~\textrm{where} ~~l \equiv e, \mu, \tau.
\label{Eq:N_decay}
\end{equation}
A representative diagram for decay of $N$ ($N\rightarrow \ell^{\mp} W^{\pm}$) is shown in Fig.~\ref{fig:pp_to_Rnl}(right plot).

\begin{figure}[t]
\includegraphics[width=5.60cm, angle =-90]{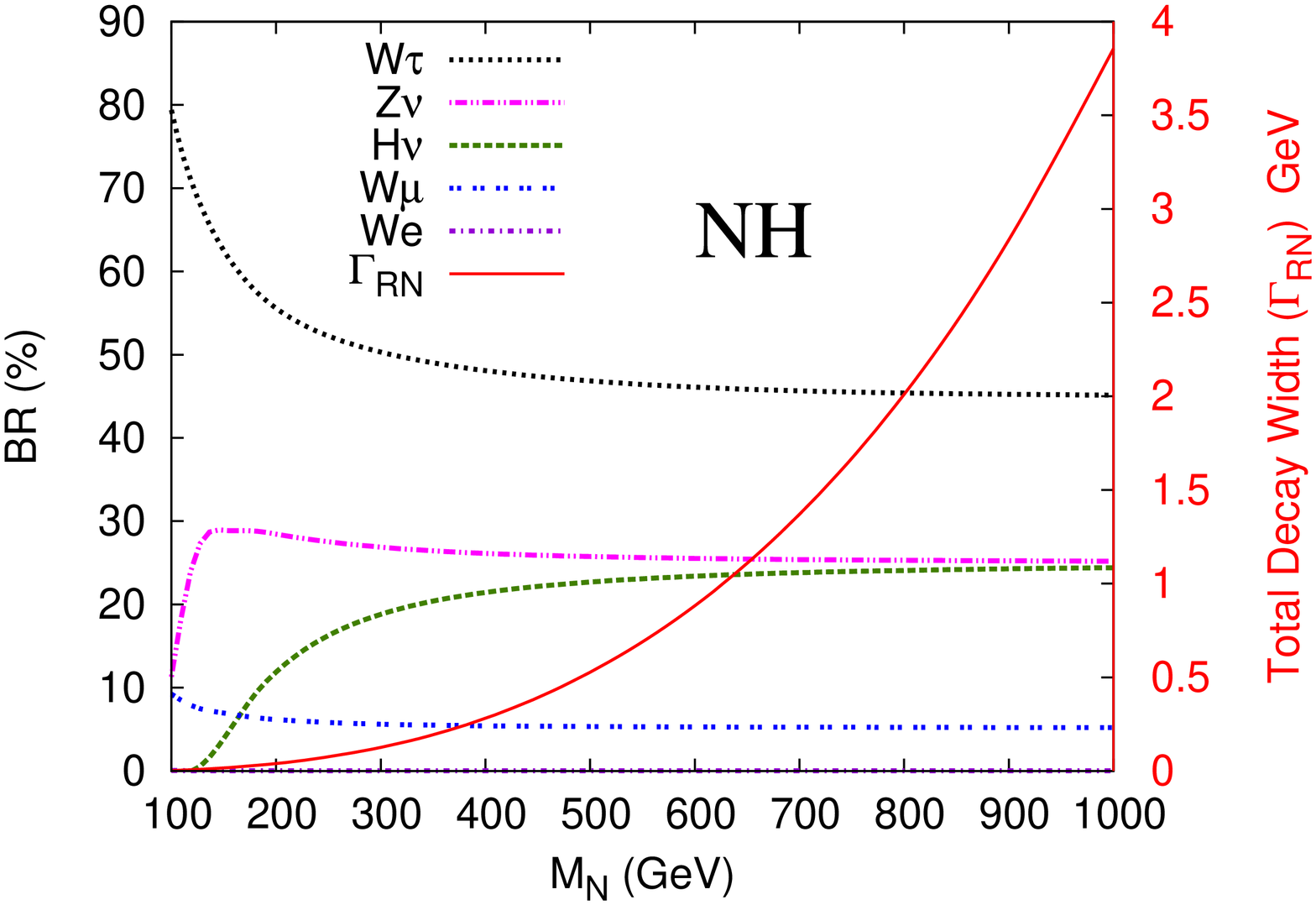} 
 \includegraphics[width=5.60cm, angle =-90]{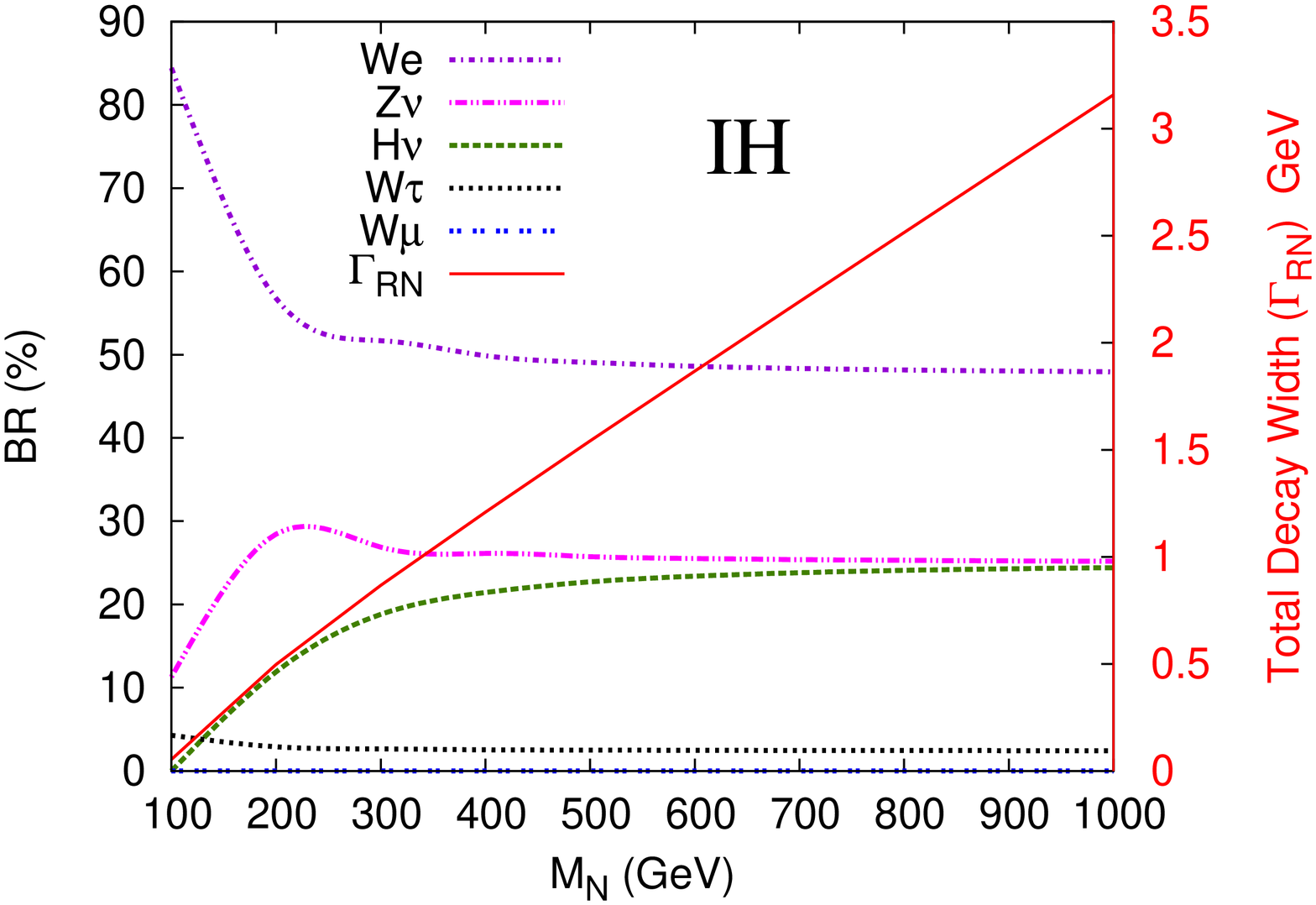} 
\caption{The decay branching ratios of the heavy neutrino ($N$) in different channels as a function 
of its mass in the case of normal hierarchy, Case-I,
(left) and inverted hierarchy (right). 
Total decay widths in each case are also demonstrated with the solid line in the same figure.}
\label{fig:RN_BR_NH}
\end{figure}

In Fig.~\ref{fig:RN_BR_NH} we present the branching ratios for these decay channels as a 
function of heavy neutrino mass $M_N$ both in the case of normal hierarchy (left) and inverted hierarchy (right).
Total decay widths in each case are also demonstrated with the solid line in each figure. 
Identifying that the charged lepton decay modes for heavy neutrino i.e. $N \to W^\pm l^\mp$ 
being the main channel for search at the hadron collider, we discuss the corresponding decay 
modes in detail for both scenarios.
The figure clearly shows that for NH, Case-I\footnote{For Case-II, although BRs to different channels 
likely to change, we do not show the corresponding plot as final production 
cross-section for both
the cases, after putting all the selection criteria, is very low for NH and 
beyond the reach of LHC at 14 TeV even 
with a luminosity of $3000~fb^{-1}$.}, heavy neutrinos mostly decay 
into tau lepton ($\tau$) and $W$ boson. On the other hand for IH, 
decay into the first generation lepton ($e$) 
possesses the maximum branching ratio. 
For NH the decay to $\mu$ is low and decay to $e$ is severely suppressed, while for IH, the decay
to $\tau$ has a lower ratio and decay to $\mu$ is negligible. The $W^{\pm}$ can have hadronic  decay modes ($W^{\pm} \to jj$) or  
leptonic decay modes ($W^{\pm} \to \l^{\pm} \nu$). The  tri-lepton  signal
 $pp \rightarrow l^{\pm} l^{\mp} \l^{\pm} \nu$ comes from the later decay mode\footnote{Evidently former decay mode leads to opposite sign dileptons (OSDL),
also suppressed by $|V_{lN}|^4$, but slightly larger compare to tri-lepton signal. However, significant irreducable backgrounds can come from $t\bar{t}$, $VV$ (with $V = W,Z $), as well as
$Z + Jets$ after vetoing dilepton invariant mass at Z-pole. Hence we are not
considering the OSDL as a signal.
Estimate of these backgrounds for OSDL can be found in \cite{Eckel:2014dza}. Note that their more specific selection criteria are not applicable for our present signal.
Similarly, OSDL through VBF is suppressed by $|V_{lN}|^4$ and is beset with large background coming from $W W$, $\tau\tau$ and $ZZ$ production at VBF\cite{Choudhury:2003hq}.}.

Other than charged lepton decay mode, $N$ can also decay to Z-boson or 
Higgs boson associated with neutrinos as listed in Eq.~\ref{Eq:N_decay}. 
The corresponding branching ratios are also shown in Fig.~\ref{fig:RN_BR_NH}.  
Note that the branching ratio for $Z\nu$ is  suppressed  
for lower values of the masses of the heavy neutrinos
essentially because of  W mass threshold. For the $H\nu$ decay mode,  
 the Higgs mass threshold suppresses the decay rate for 
lower values of $M_N \sim 100$ GeV.  
However, as $M_N$ increases these branching ratios increase to retain a
$\sim$ 25\% level.
Both these channels can contribute to the tri-lepton signal via leptonic decays and we
have considered their contributions in our simulation. 
However since we will apply Z-veto (to minimize the SM background), the 
contribution coming from $Z\nu$ decay mode will be suppressed after final event selection. 

As lepton Yukawa is small, the $H\nu$ mode is also not going to contribute to our signal even for
higher values of $M_N$.

\section{Simulation and Results}\label{sec:simulation_result}

We have implemented the model in {\tt FeynRules}~\cite{Christensen:2008py} and generated the Feynman rules compatible 
with {\tt MadGraph5}~\cite{Alwall:2011uj}. 
After generating Les Houches Event (LHE)~\cite{Alwall:2006yp} file from {\tt MadGraph}, we have  passed that to 
{\tt PYTHIA6}~\cite{Sjostrand:2006za} for showering and hadronization. 

\subsection{Selection criteria}\label{subsec:cuts}
To get enhancement in signal over background, we use the following selection criteria~\cite{Bambhaniya:2013yca,Mukhopadhyaya:2010qf}:

\begin{itemize}
\item   Identification criteria of a lepton: pseudorapidity $|\eta_{\ell}| < 2.5$ and ${p_T}_{\ell}  >$  20 GeV have been used.  
\item   Detector efficiency for leptons~\cite{Aad:2009wy,Ball:2007zza}: 
\begin{itemize}
\item   For electron (either $e^-$ or $e^+$) detector efficiency is 0.7 ($70\%$);
\item   For muon (either $\mu^-$ or $\mu^+$) detector efficiency is 0.9 ($90\%$).
\end{itemize}
\item Smearing\footnote{Choice of corresponding $\eta$ dependent parameters is similar to one of our earlier work~\cite{Bambhaniya:2013yca}.} of electron energy and muon $p_T$ are incorporated. 
\item Lepton-lepton separation:  for this  $\Delta R_{ll} \ge 0.2$  is used\footnote{Here $\Delta R_{ij} = \sqrt{(\eta_i -\eta_j)^2 -(\phi_i - \phi_j)^2}$ quantifies the 
separation between particles $i$ and $j$ in the pseudorapidity($\eta$)-azimuth($\phi$) plane.} (due to detector resolution 
of leptons).
\item Lepton-photon separation:  this is taken as 
$\Delta R_{l\gamma} \ge 0.2$ with all the photons having ${p_T}_\gamma > 10$ GeV.
\item Lepton-jet separation: The separation of a lepton with all the jets 
is set at $\Delta R_{lj} \ge 0.4$; 
otherwise that lepton is not counted as lepton. 
Jets are constructed from hadrons using {\tt PYCELL} within the {\tt PYTHIA}.
\item Hadronic activity cut: This cut is applied to take only pure kind of leptons that have very less hadronic 
activity around them. The hadronic activity within the cone of radius 0.2 around the lepton should be small, 
$\frac{\sum p_{T_{hadron}}}{p_{T_l}} \le 0.2$. 
\item Hard $p_T$ cuts used are:  ${p_T}_{l_1}>30$ GeV, ${p_T}_{l_2}>30$ GeV and ${p_T}_{l_3}>20$ GeV. 
\item Missing $p_T$ cut: Due to the presence of 
neutrino, a missing  $p_T$ cut ($ > 30$ GeV) is applied.
\item Z-veto\footnote{Same flavored but opposite sign lepton pair invariant mass $m_{\ell_1
\ell_2}$ must be sufficiently away from $Z$ mass, such that, typically, 
$|m_{\ell_1\ell_2} - M_{Z}| \geq 6 \Gamma_{Z} \sim 15$ GeV.} is applied to suppress the SM background. 

\item VBF cuts~\cite{Rainwater:1999gg,Yazgan:2007cd}:
\begin{itemize}
\item Central jet veto is also applied, in which we consider any jet with ${E_T}_3 > 20$ 
GeV and compute the rapidity with respect to
the average of the two forward jets: $ \eta_0 = \eta_3 - (\eta_1 + \eta_2)/2 $.
We veto the event if $|\eta_0|  < 2$. Central jet veto is applied to suppress the QCD background substantially.
\item Charged leptons need to fall in between the rapidities of two forward tagging jets i.e. 
$\eta_{j,min}<\eta_{\ell}<\eta_{j,max}$.
\item $p_T$ of jets: ${p_T}_{j_1,j_2} > 20$ GeV.
\item Invariant mass of jets: $M_{j_1j_2} > 600$ GeV.
\item Pseudorapidity of jets: $\eta_{j_1}.\eta_{j_2}<0$ and $|\eta_{j_1}-\eta_{j_2}|>4$. 
Demanding both the tagged jets in opposite hemisphere and a large rapidity separation among them significantly reduces the BG for VBF.
\end{itemize}

\end{itemize}

\subsection{Background}

\subsubsection{For s-channel signal}

To calculate the SM background we consider all  channels that can produce or mimic 
the tri-lepton production with missing $P_T$. We closely follow
the reference~\cite{Bambhaniya:2013wza,Bambhaniya:2013yca}
where similar background analysis was done with the event selection 
criteria listed as above  except the cuts related to the VBF. 
Events are generated using {\tt ALPGEN}~\cite{Mangano:2002ea} 
for the processes coming from 
$t \bar{t}$, $t \bar{t}(Z/\gamma^*)$, $t \bar{t}W^{\pm}$, $W^{\pm}(Z/\gamma^*)$, $(Z/\gamma^*)(Z/\gamma^*)$ 
at the parton level and passed into {\tt PYTHIA}. As expected $t \bar{t}$ and $W^{\pm}(Z/\gamma^*)$ contribute dominantly. 
These and other SM backgrounds are listed in Table~\ref{table:background_events_s-ch}. 
For each process we classify the tri-lepton signals into four different flavor combinations and 
compute the cross section in each case along with the total contribution.

\begin{table}[t]
\begin{tabular}{|c|c|c|}
\hline
Process & \multicolumn{2}{c|} {Cross section ($fb$)}\\
\cline{2-3}
& ~~$\ell\ell\ell$~~ &$eee$ \hspace{0.8cm}    $ee\mu$  \hspace{1cm} ~~$e\mu\mu$ \hspace{1cm} $\mu\mu\mu$\\\hline
$t \bar{t}$                   &18.972  &1.1383  ~~~ 7.0831   ~~~~ 8.2214  ~~~~ 2.5297  \\
\hline
$W^{\pm}(Z/\gamma^*)$         &10.832  &0.0677  ~~~ 0.1311   ~~~~ 5.9891  ~~~~ 4.6440  \\
\hline
$(Z/\gamma^*)(Z/\gamma^*)$    &1.175   &0.0734~~~ 0.0525   ~~~~ 0.6400  ~~~~   0.4090 \\
\hline
$t \bar{t}(Z/\gamma^*)$       &1.103   &0.0429  ~~~ 0.1329   ~~~~ 0.4997  ~~~~ 0.4275  \\
\hline
$t \bar{t}W^{\pm}$            &0.639   &0.0328  ~~~ 0.2655   ~~~~ 0.2424  ~~~~ 0.0983  \\
\hline
{\bf TOTAL}      &  {\bf ~~32.721~~}  & {\bf 1.3551 ~ 7.6651 ~~ 15.5926 ~ ~8.1085}  \\
\hline
\end{tabular}
\caption{Dominant Standard Model background cross sections contributing 
to tri-lepton and missing transverse energy. These are calculated satisfying all the cuts (except VBF cuts) 
for the 14 TeV LHC. For each process we also classify the tri-lepton background into four different flavor 
combinations and present the cross section in each case along with the total contribution.}
\label{table:background_events_s-ch}
\end{table}

\subsubsection{For VBF signal}

Tri-lepton signal with missing $P_T$ and two forward jets
in VBF can be faked by different SM backgrounds. 
Processes like $t\bar{t}$ would produce b-jets and mostly effective in central region. 
Vetoing on jet activities in central region can eliminate most of the non-VBF type SM processes. 
However most important irreducible background comes from $W^{\pm} Z$ and $Z Z$ together with two 
extra forward jets once the gauge bosons decay leptonically. These processes can construct dominant 
SM background for the VBF production of $3\ell+\MET$ since they includes the typical VBF topology 
and hence can easily pass the central jet veto criteria. These backgrounds are 
calculated\footnote{Next to leading order QCD corrections are available in~\cite{Bozzi:2007ur, Jager:2006cp}.} 
using {\tt MadGraph5} and {\tt PYTHIA6}. In the Table~\ref{table:background_events} the dominant background 
cross sections after satisfying all the cuts including VBF cuts at 14 TeV LHC is tabulated. Like the case of s-channel 
backgrounds, for each process we also classify the tri-lepton signals into four different flavor 
combinations and compute the cross section in each case as well as the total contribution.

\begin{table}[t]
\begin{tabular}{|c|c|c|}
\hline
Process & \multicolumn{2}{c|} {Cross section ($fb$)}\\
\cline{2-3}
& ~~$\ell\ell\ell$~~ &eee \hspace{0.8cm}    $ee\mu$  \hspace{1cm} ~~$e\mu\mu$ \hspace{1cm} $\mu\mu\mu$\\\hline
$W^{+}Zjj$ &0.04068  &0.00073  ~~~ 0.00105   ~~~~ 0.02157  ~~~~ 0.01734  \\
\hline
$W^{-}Zjj$ &0.01923 &0.00038  ~~~ 0.00055  ~~~~ 0.00994 ~~~~ 0.00836 \\
\hline
$ZZjj$     &0.00094 &0.00002 ~~~ 0.00002 ~~~~ 0.00066 ~~~~ 0.00024 \\
\hline
{\bf TOTAL}      &  {\bf ~~0.06085~~}  & {\bf 0.00113 ~ 0.00162 ~~ 0.03216 ~ ~0.02594}  \\
\hline
\end{tabular}
\caption{Dominant Standard Model background cross section contributing 
to tri-lepton and missing transverse energy associated with two forward jets. These are calculated satisfying all the cuts 
including VBF cuts for the 14 TeV LHC. Cross sections of four different flavor combinations as well as the total cross
section are listed.}
\label{table:background_events}
\end{table}

\subsection{Signal}

Earlier in section.~\ref{sec:phenomenology} we have presented the total heavy neutrino 
production cross sections for different light neutrino hierarchy with basic 
selection criteria.  
The crossection for NH scenario was found to be much lower than the 
IH scenario for $s$-channel. 
The brancing ratios for  decays of N to final states with $\mu$ and $e$ are
also very small for NH.  
Therefore we will concentrate only on IH scenario henceforth. 
For this we consider both 
s-channel and VBF process.  Although the VBF cross-section for IH is lower 
or comparebale to s-channel cross-section for NH for lower values of 
$M_N$, the background for VBF procceses are much smaller. Hence we study 
this channel also for IH. 
In this section we consider all leptonic 
decay modes of heavy neutrinos for a benchmark mass of $M_N$ at 100 GeV  with the cuts discussed in section.~\ref{subsec:cuts}.

\subsubsection{Signal for s-channel }
The signal coming from decay of heavy neutrinos 

$$p p \rightarrow \ell^{\pm} N  \rightarrow \ell^{\pm} ( \ell^{\mp} W^{\pm}) 
\rightarrow \ell^{\pm}\ell^{\mp}\ell^{\pm}+ \MET,\hspace{0.3cm} \textrm{ \;where $\ell \equiv e\,, \mu$}.$$

Table~\ref{table:signal_Schannel}~~ lists the final tri-lepton signal cross section  through s-channel heavy 
neutrino production at 14 TeV LHC for the benchmark point $M_N = 100$ GeV incorporating all event selection 
criteria except VBF cuts as described earlier. 
The total contribution from the light leptons as well as the contributions 
from the four different flavor combinations are presented.

\begin{table}[t]
\begin{tabular}{|c|c|c|}
\hline
Hierarchy  & \multicolumn{2}{c|} {Cross section ($fb$)}\\
\cline{2-3}
& ~~$\ell\ell\ell$~~ &~~~~~eee ~~~~~~~~    $ee\mu$  ~~~~~~  $e\mu\mu$ ~~~~   $\mu\mu\mu$~\\
\hline
IH & ~27.07~  &~10.297 ~~ 16.314 ~~ 0.459 ~~~~ 0.0  \\
\hline
\end{tabular}
\caption{Cross section for IH case. Final tri-lepton signal cross section through s-channel heavy neutrino production at the 14 TeV LHC for the
benchmark point $M_N = 100$ GeV including all event selection cuts except VBF cuts. 
We classify the tri-lepton signals into four 
different flavor combinations and present the cross section in each case along with the total light lepton contribution. }
\label{table:signal_Schannel}
\end{table}

 As we can see from the Table~\ref{table:signal_Schannel} cross section in terms of flavors has the ordering:
$e e\mu > eee > e \mu\mu > \mu\mu\mu $.
We can understand this in the following way.
There are total 8 possibilities which can  produce $\ell\ell\ell$ events.  
There is only one way to produce
$\mu\mu\mu$ and $eee$ final states.
However, there are three possible ways to get the  $e e\mu$ channel
depending on which one of $\ell_i$'s in figure~\ref{fig:pp_to_Rnl} is associated with 
$e$ and $\mu$.  Similarly for the  $e\mu\mu$ final state also we get 3 possibilities. 
The amplitude for $eee$ channel $\sim V_{eN}^4$; 
the  $ee\mu$ channel goes as  $\sim V_{eN}^2 + 2 V_{eN} V_{\mu N}$;
the $e\mu\mu$ channel goes as
$\sim V_{\mu N}^2 + 2 V_{eN} V_{\mu N}$ while the $\mu \mu \mu$ channel 
as $\sim V_{\mu N}^2$.
Since $V_{e N} >> V_{\mu N}$, the $eee$ and $ee\mu$ cross sections are much larger whereas $\mu\mu\mu$ cross section is negligible.
$ee\mu$ crossection is higher than the $eee$ crossection because of higher muon efficiency in the detector, whereas 
the small $e\mu\mu$ crossection is due to a very tiny value of $V_{\mu N}$. 

One can also compute the ratios of events with different flavour 
compositions in which some of the common systematic uncertainties can 
get cancelled. For example  
$ee\mu/{eee} \sim \epsilon$ where $\epsilon$ denotes the relative efficiency of
detection of muon over electron,
${ee\mu}/{\mu\mu\mu} \sim  \epsilon V_{eN}^4/V_{\mu N}^4$; 
${eee}/{e\mu\mu} \sim \epsilon^2 V_{eN}^4/V_{\mu N}^4$ etc. 
Since  for a fixed $y_\nu$, which in turn implies specific 
values for phases, the variation of the light-heavy mixing angles are
not very much with oscillation parameters, these ratios vary within a very 
narrow range\footnote{Note that, the allowed magnitude of mixings are as following: For a fixed value of $y_\nu\,(=0.4)$, $\alpha\,(=3\,\pi/4)$, $\delta\,(=0)$ 
and $M_N\,(=100)$ GeV, the magnitude 
of $|V_{e N}|$ and $|V_{\mu N}|$ vary in a very small range for $3\,\sigma$ variation of oscillation parameters; 
$|V_{e N}| = 0.471 - 0.484$, $|V_{\mu N}| = 1.236\times 10^{-4} - 1.272\times 10^{-4}$. However, $|V_{\tau N}|$ varies
little higher; $|V_{\tau N}| = 0.092 - 0.147$. Since we are considering  modes 
involving only $e$ and $\mu$,  the crossections 
are likely to vary by a small amount for different set of oscillation 
parameters.} and hence can be used to test the model.  
Ofcourse for different phase choices a different $y_\nu$ and hence different 
predictions can be obtained. However, a smaller value in $y_\nu$ would result 
in a lower event rate and hence it would be difficult to test at the LHC.

\subsubsection{Signal for VBF }

In this section we present the results for the case where
$N$ is produced by VBF: 
$$p p \rightarrow \ell^{\pm} N j j \rightarrow \ell^{\pm} (\ell^{\mp}W^{\pm}) j j 
\rightarrow \ell^{\pm}\ell^{\mp}\ell^{\pm}+ \MET + j j \textrm{(forward jets)},\hspace{0.3cm}
\textrm{ \;where $\ell \equiv e\,, \mu$}.$$

In Table~\ref{table:signal_background}  we present the final tri-lepton signal cross sections through VBF production 
of heavy neutrinos at the 14 TeV LHC for the benchmark point $M_N = 100$ GeV, after including all cuts.
Here we have only shown the case of inverted hierarchy and signal is 
found to be quite small.
Although VBF backgrounds are small, the tiny production cross sections are 
insufficient for giving any signal with integrated luminosity of 300 $fb^{-1}$. 
Some indications from VBF can 
appear only at the HL-LHC (3000 $fb^{-1}$).  
However, $5\sigma$  significance can not be reached even for $M_N=100$ GeV.

\begin{table}[t]
\begin{tabular}{|c|c|c|}
\hline
Hierarchy & \multicolumn{2}{c|} {Cross section ($fb$)}\\
\cline{2-3}
& ~~$\ell\ell\ell$~~ &~~~~~eee \hspace{1.3cm}    ~~~~$ee\mu$  \hspace{1.3cm} ~~~~~$e\mu\mu$ \hspace{1.3cm} $\mu\mu\mu$~~~~\\
\hline
IH & ~0.018068~  &$7.09\times10^{-3} $\hspace{0.6cm} $1.06\times10^{-2} $\hspace{0.6cm} $4.06\times10^{-4} $\hspace{0.6cm} $0.00$  \\
\hline
\end{tabular}
\caption{Final tri-lepton Signal through VBF production of heavy neutrino for the benchmark point $M_N = 100$ GeV at 14 TeV LHC for IH after all event selection cuts. }
\label{table:signal_background}
\end{table}

\section{Discovery potential}\label{sec:discovery_pot}

\begin{figure}[t]
\includegraphics[width=4.9cm, angle =-90]{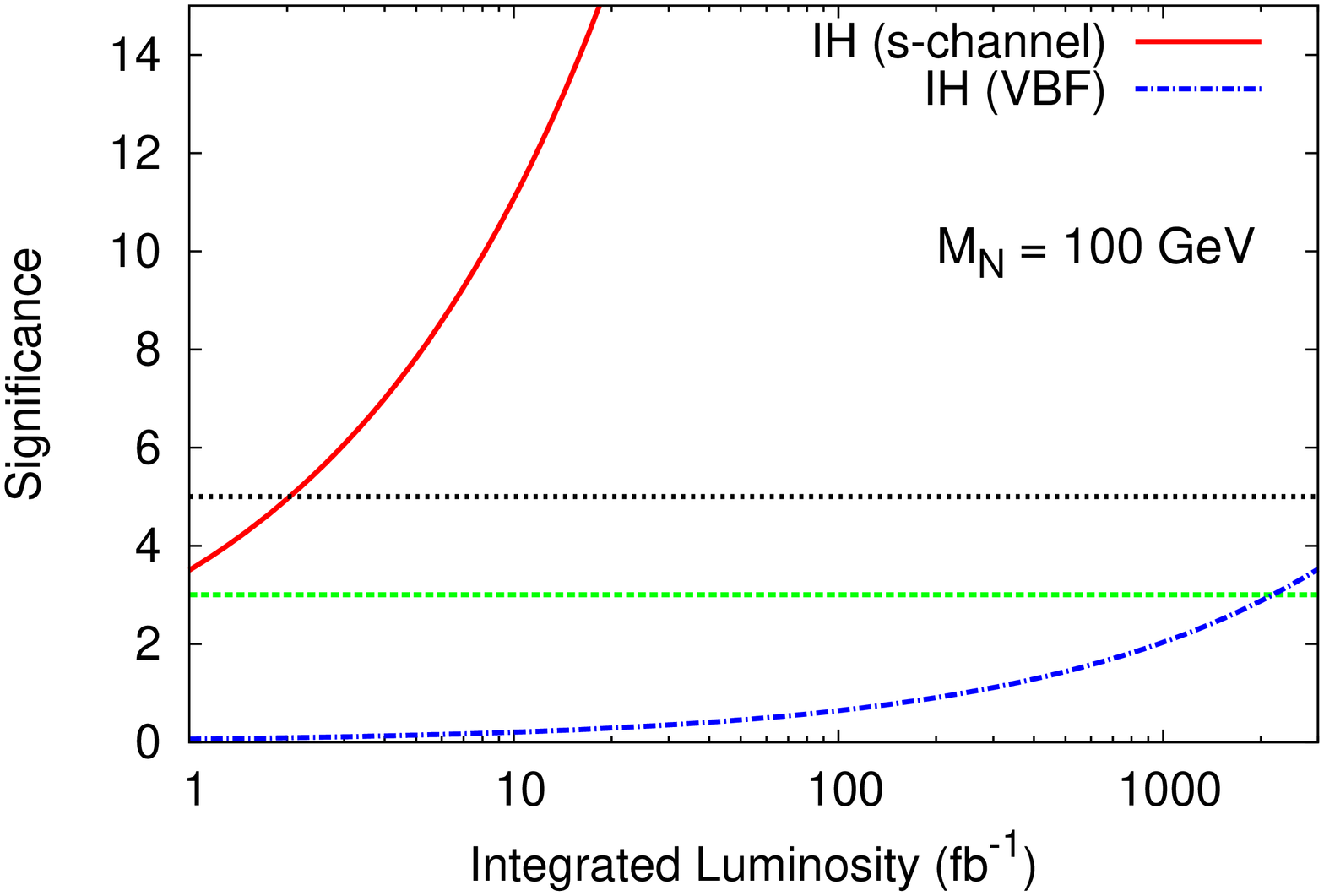} 
\includegraphics[width=4.9cm, angle =-90]{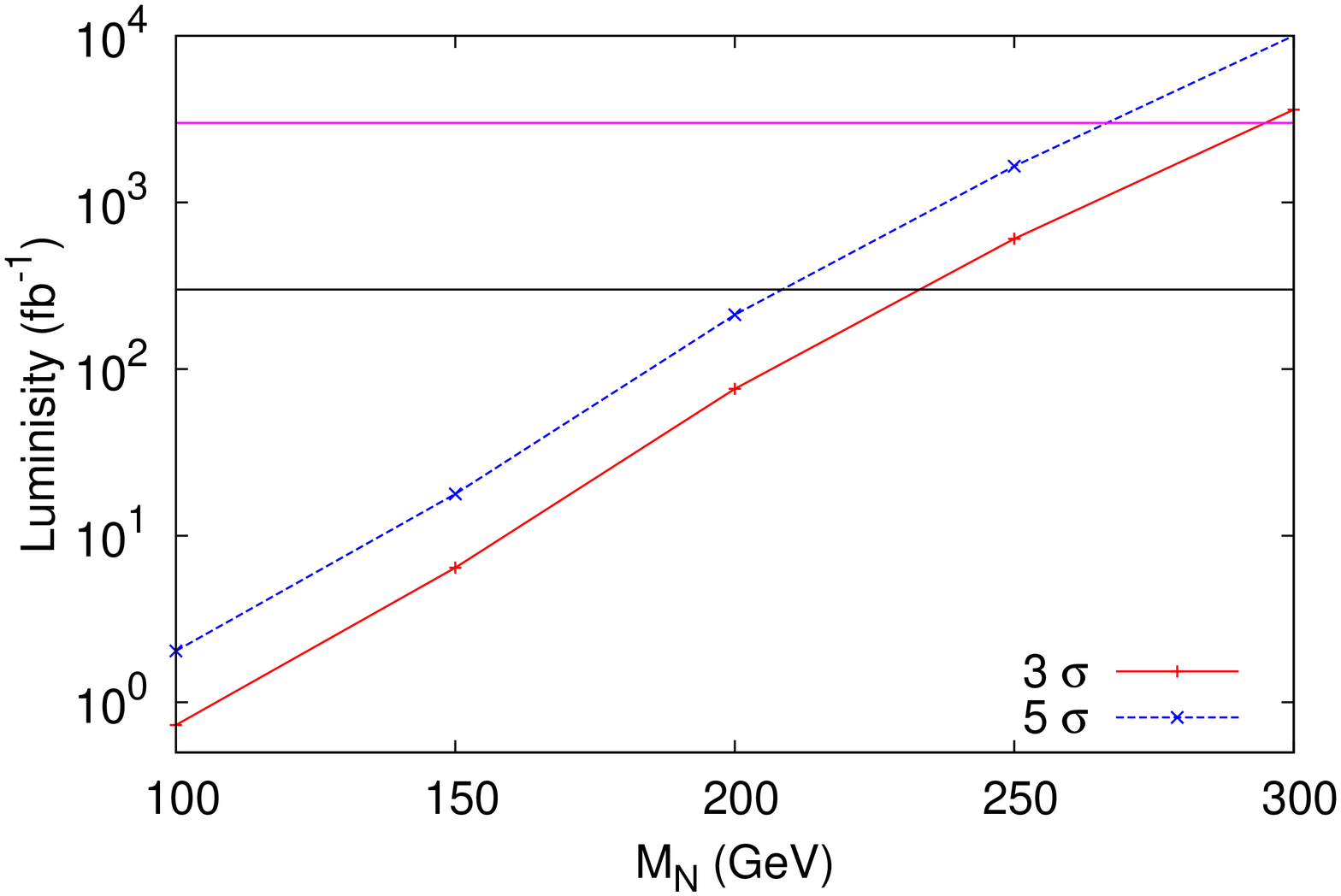} 
\caption{(Left) The variation of significance ${S}/{\sqrt{S+B}}$ for the
s-channel 
production signal for benchmark point $M_N = 100$ GeV 
with the integrated luminosity available for the low luminosity option at 14 TeV LHC.
Black-dotted (green-dashed) line parallel to the x-axis
represents 5$\sigma$ (3$\sigma$) significance.
(Right) The lines for $3\sigma$(red) and $5\sigma$(blue) 
significance in terms of heavy neutrino mass and integrated luminosity.
With $300 fb^{-1}$ luminosity at LHC14 the heavy neutrino mass in this model 
can be probed up to $\sim 210$($230$) GeV with $\sim$5$\sigma$ (3$\sigma$) 
significance.
For very high luminosity of $3000 fb^{-1}$ this can reach up to $\sim 270(295)$ GeV.
 }
\label{plot:significance}
\end{figure}

After numerical computation of all necessary signals and backgrounds, results are better represented in terms of 
significance, defined as ${S}/{\sqrt{S+B}}$, where $S (B) = \mathcal{L} \sigma_{S(B)}$. Here $\mathcal{L}$ being
integrated luminosity available for the collider at certain machine energy and $\sigma_{S(B)}$ is the final cross section
after all event selection, for given parameters like heavy neutrino mass and corresponding allowed couplings.
Fig.~\ref{plot:significance} (Left) demonstrates the expected significance coming from 
s-channel production of heavy Dirac  neutrino of mass $100$ GeV as a function of integrated luminosity at 14 TeV LHC. 
In the figure 
black-dotted (green-dashed) line shows 5$\sigma$ (3$\sigma$) significance. 
From the figure it is clear that
for the case of s-channel signal in the IH scenario, 3$\sigma$(5$\sigma$) significance can be achieved within the integrated luminosity $\sim$ 0.73(2.03) $fb^{-1}$. In the case of 
VBF channel 3$\sigma$ significance can be achieved with 2175 $fb^{-1}$ luminosity, while 5$\sigma$ significance is not achievable within 3000 $fb^{-1}$ luminosity which is planned for the HL-LHC.

 Fig.~\ref{plot:significance} (Right) shows the lines for $3\sigma$(red) and $5\sigma$(blue) significance in terms of heavy neutrino mass and integrated luminosity.
 With $300 fb^{-1}$ luminosity at LHC14 the heavy neutrino mass in this model can be probed up to $\sim 210$($230$) GeV with $\sim$5$\sigma$ (3$\sigma$) significance.
For very high luminosity of $3000 fb^{-1}$ this can reach up to $\sim 270(295)$ GeV. For VBF signal, since $M_N = 100$ GeV itself requires a very large integrated luminosity; higher values of $M_N$ are not 
possible to explore.

\section{Summary and Conclusion}\label{sec:conclusion}

In this work we have considered TeV scale minimal linear seesaw model which 
generates correct order of light neutrino masses and has sizable light-heavy mixing to produce heavy neutrinos at colliders like LHC. 
One of the important features of this model is that it can be fully reconstructible from oscillation data excepting 
an overall factor $y_\nu$ characterizing the Dirac Yukawa matrix. However this parameter 
gets constrained by LFV and vacuum meta-stability bounds. The neutral fermion mass spectrum of this model 
consists of one massless, two light and two heavy neutrinos. 

We have studied the collider phenomenology of TeV scale linear seesaw at 14 TeV LHC.
The heavy neutrinos in this model can be 
dominantly  produced through the  s-channel. 
In a leading order calculation,
subsequent decay of these leads to characteristic tri-lepton signal with 
missing $p_T$. We also consider the production of heavy neutrinos 
through the VBF process.  
The signal for this is tri-leptons
with additional two forward jets which can be tagged. 
Both these signals as well as  SM backgrounds have been estimated with realistic simulations using 
{\tt MadGraph} and {\tt PYTHIA}. 

We found that s-channel  tri-lepton production process have potential to be 
discovered at the LHC for IH scenario. However due to severe constraint on the 
light-heavy mixing coming form LFV in the 
case of NH scenario, both $s$-channel and VBF can not be probed at the $14$ TeV LHC with proposed luminosity. For a benchmark point with a heavy neutrino mass $M_N =100$ GeV, $3\sigma$ 
significance can be achieved with integrated luminosity of $\sim 0.73(2175)~fb^{-1}$ for $s$-channel(VBF) signal in the IH scenario. 
$5\sigma$ significance can be reached for $s$-channel signal with a integrated luminosity of $\sim 2~fb^{-1}$, however for VBF signal 
the required luminosity is $\sim 6042~fb^{-1}$, which is beyond the reach of projected luminosity at the LHC.
 Discovery reach in the tri-lepton channel can be achieved upto the heavy neutrino mass of $\sim 210$($230$) GeV with $\sim$5$\sigma$ (3$\sigma$) 
significance at the low luminosity ($300~fb^{-1}$) option of 14 TeV LHC. 
 In the high luminosity ($3000~fb^{-1}$) search, reach is upto $\sim 270(295)$ GeV.
 Whereas, VBF channel can only reach upto $\sim 3\sigma$ for $M_N$ at 100 GeV. 
 Our analysis uses values for the 
elements, $V_{lN}$ of the light-heavy mixing matrix, 
which are consistent with the constraints coming from vacuum metastability 
and LFV.  
Any freedom of choosing larger values 
({\sl e.g.} $\sim \mathcal{O}(1)$) for these parameters 
can extend the discovery limit by a very significant amount. 
With the constraints used in this work, for $V_{lN}$,  a detectable 
tri-lepton signal can only be obtained for the inverted hierarchical scenario with particular choices of phases leading to large $y_{\nu}$. 
One can also compute the ratios of events with different flavour  
compositions which are proportional to the elements $V_{lN}$.  
They vary only within a narrow range with the $3\sigma$ variations of oscillation
parameters and thus the model has very definite predicions for these ratios.

\bibliographystyle{plain}
\bibliographystyle{apsrev4-1}
\bibliography{bibliography}

\end{document}